\documentclass[11pt]{article}

\usepackage[a4paper,margin=1in]{geometry}
\usepackage{amsmath,amssymb,amsthm}
\usepackage{graphicx}
\usepackage{hyperref}
\usepackage{mathtools}
\usepackage{braket}

\usepackage{authblk}

\usepackage{amsmath}
\usepackage{amssymb}
\usepackage{graphicx,color,slashed,braket}
\usepackage{bbm}
\usepackage{ifpdf}
\usepackage{comment}
\usepackage{soul}

\numberwithin{equation}{section}

\usepackage{dsfont}

\usepackage[absolute,overlay]{textpos}

\usepackage{booktabs}
\usepackage{multirow}
\usepackage{makecell}
\usepackage{float}
\usepackage{verbatim}
\usepackage{caption}
\usepackage{cancel}
\usepackage{enumerate}
\usepackage{tcolorbox}
\usepackage[font=small,labelfont=bf]{caption}
\usepackage[numbers,sort&compress]{natbib}

\usepackage{color}
\definecolor{dark-gray}{gray}{0.20}
\definecolor{gray}{gray}{0.30}
\definecolor{light-gray}{gray}{0.80}
\definecolor{dark-red}{rgb}{0.7,0,0}
\definecolor{dark-green}{rgb}{0.1,0.4,0}
\definecolor{dark-blue}{rgb}{0.3,0.3,0.7}
\definecolor{light-blue}{rgb}{0.8,0.8,1}

\usepackage{tikz}
\usetikzlibrary{calc}

\usepackage{hyperref}
\usepackage{setspace}
\hypersetup{
	colorlinks=true,
	linkcolor=dark-blue,
	citecolor=dark-red,
	urlcolor=dark-green,
	linktoc=page
}
\newcommand{\be}{\begin{equation}}
\newcommand{\ee}{\end{equation}}

\def\be{\begin{equation}}
\def\ee{\end{equation}}
\def\bea{\begin{eqnarray}}
\def\eea{\end{eqnarray}}

\newcommand{\e}{\mathrm{e}}

\newcommand{\dd}{\mathrm{d}}

\newcommand{\Mpl}{M_\text{Pl}}

\renewcommand{\Im}{\text{Im}\,}
\renewcommand{\Re}{\text{Re}\,}

\allowdisplaybreaks

\title{\bf 
$\mathcal{N}=1$ spectra, cubic couplings\\ and the rigid fate of DGKT}

\author[1]{Filippo Revello}
\author[2]{Vincent Van Hemelryck}

\affil[1]{Instituut voor Theoretische Fysica and Leuven Gravity Institute, KU Leuven, Celestijnenlaan 200D, B-3001 Leuven, Belgium\newline}
\affil[2]{Department of Physics and Astronomy, Uppsala University, Box 524, SE-75120, Uppsala, Sweden}
\date{}

\begin{document}
\begin{textblock*}{4cm}(13.7cm,2cm)
\raggedleft
\small UUITP-16/26
\end{textblock*}
\maketitle
\vspace{1cm}
\begin{abstract}\noindent
    We show that in the DGKT scenario on a generic Calabi--Yau three-fold, a recently proposed holographic constraint on cubic couplings is satisfied if and only if the Calabi--Yau is rigid, i.e.~when $h^{2,1}=0$. 
    More generally, we illustrate how in 4d $\mathcal{N}=1$ supergravity, extremal cubic couplings are determined by the third derivatives of the real, K\"ahler-invariant superpotential, while the eigenvalues of its Hessian compute the conformal dimensions of the dual scalar operators. These results extend more broadly beyond 4d $\mathcal{N}=1$ supergravity. Applying them to supersymmetric DGKT vacua, we prove that extremal cubic couplings always vanish in the K\"ahler + universal CS/dilaton sector, whereas non-vanishing \mbox{(super-)extremal} couplings are always present in the complex structure sector. It follows that the holographic constraint is satisfied in DGKT if and only if the Calabi--Yau three-fold is rigid with $h^{2,1}=0$.
\end{abstract}

\newpage
\tableofcontents
\vspace{0.5cm}
\hrule

\section{Introduction}\label{sec:intro}

The existence of a ``cosmic hierarchy" between the size of the observed universe and that of the (unobserved) extra dimensions is a fundamental question that must be addressed in any attempt to reconcile (geometric) string compactifications with the real world. In technical terms, this is known as the problem of scale separation -- between the scale of the ``largest" extra dimension (or inverse mass of the lightest Kaluza-Klein mode) and the four-dimensional Hubble scale. Regardless of its phenomenological implications, this puzzle can also be turned into an important and abstract question in the context of holography. In particular, one might ask whether there exist any instances of the AdS/CFT correspondence involving the product of an AdS space and a compact manifold whose radii are all parametrically smaller than the AdS radius \cite{Polchinski:2009ch,Alday:2019qrf}. Along the same lines, the existence of scale-separated AdS vacua would also imply the existence of bizarre CFTs with a large gap in the spectrum of scalar primaries, of which no examples are currently known \cite{Collins:2022nux}.

However, fully-established scale-separated constructions in string theory remain elusive, as all known top-down examples from string compactifications involve some debated approximations. For instance, there has been criticism on the constructions involving only classical ingredients, pioneered by refs.~\cite{DeWolfe:2005uu,Camara:2005dc}, because of the presence of orientifold planes. This criticism has been mostly refuted by refs.~\cite{Baines:2020dmu,Junghans:2020acz, Marchesano:2020qvg,Junghans:2023yue}. Nevertheless, further testing the validity of scale-separated AdS vacua remains one of the central challenges in string phenomenology and the swampland program (see \cite{Coudarchet:2023mfs} for a review).

From this point of view, holography offers a promising avenue to look for non-trivial consistency checks that bypass specific issues in the explicit constructions. Although complete holographic duals for these scale-separated vacua are not known \cite{Aharony:2008wz}, one can still ask whether putative dual CFTs satisfy general constraints that any consistent holographic theory must obey \cite{El-Showk:2011yvt} (see also \cite{Bobev:2023dwx,Perlmutter:2024noo}). In particular, the characteristic gap in the spectrum of holographic CFTs allows one to isolate a low-lying sub-sector where conditions on the spectrum and on operator couplings can be tested directly in the AdS effective field theory, without requiring a full construction. Bootstrap constraints, which have proven extremely useful to carve out the space of allowed CFTs, are unfortunately satisfied automatically at tree-level in a holographic setting \cite{Heemskerk:2009pn} (see \cite{Conlon:2018vov,Conlon:2020wmc} for more speculative attempts in this direction). However, it was recently realised in \cite{Bobev:2025yxp} that if the spectrum obeys a certain \emph{(super-)extremality} condition there does exist a non-trivial condition on the cubic couplings of the theory. \textit{Extremal} arrangements of scalar operators occur when the sum of the (tree-level) scaling dimensions of two scalar operators equals the scaling dimension of a third (minus an even integer for \textit{super-extremal} arrangements), and if this happens, then the corresponding three-point function must vanish if one insists on large~$c$ factorisation.\footnote{This is true for a proper EFT with a single cutoff, as in the case of genuine, scale-separated AdS vacua. Relaxing this assumption, counterexamples are possible \cite{Chester:2025wti,Chester:2025jxg} (see also \cite{Castro:2024cmf}).} While extremal arrangements might seem to involve a certain degree of fine-tuning, they are surprisingly common in examples of putative CFT duals to scale-separated scenarios. This is due to the fact that in a wide class of flux compactifications, including the DGKT--CFI scenario and various AdS$_3$ and AdS$_4$ vacua, the dimensions of scalar operators are integer \cite{Arboleya:2024vnp,Farakos:2025bwf, VanHemelryck:2025qok,Conlon:2021cjk,Apers:2022zjx,Apers:2022tfm,Quirant:2022fpn,Ning:2022zqx,Plauschinn:2022ztd,Andriot:2023fss,Arboleya:2025ocb,Arboleya:2025jko}. This constraint can then be explicitly checked in concrete compactifications of this type. It offers a sharp diagnostic for the consistency of AdS vacua, which is completely independent of the procedure used to construct them.

In \cite{Bobev:2025yxp}, it was shown that the constraint is non-trivially satisfied in the original realisation of DGKT, based on the orbifold $\mathbb{T}^6/ \mathbb{Z}_3 \times\mathbb{Z}_3$. This is to be contrasted with our more recent work \cite{Revello:2026eqp}, where we found that the (super-)extremal cubic couplings of the type IIB AdS$_3$ vacua of ref.~\cite{VanHemelryck:2025qok} and the type IIA AdS$_3$ vacua of ref.~\cite{Farakos:2025bwf} -- which arise as compactifications on $\mathbb{Z}_2 \times\mathbb{Z}_2 \times\mathbb{Z}_2$ orbifolds of twisted tori with $G_2$-structure -- do not vanish. However, we identified different orbifold groups that project the offending scalars out, such that all \mbox{(super-)extremal} cubic couplings vanish. Similarly, we showed that the DGKT-CFI scenario \cite{DeWolfe:2005uu,Camara:2005dc}, compactifying massive type IIA string theory on a $\mathbb{Z}_2 \times\mathbb{Z}_2$ orbifold of the six-torus, also has (super-)extremal cubic couplings that are not vanishing, contrary to the $\mathbb{Z}_3 \times\mathbb{Z}_3$ orbifold. In DGKT, non-vanishing cubic couplings always arise from complex structure moduli, whereas the latter has no such moduli. Moreover, it was recently verified that all the orbifolds with $h^{2,1}=0$ appearing in the classification of \cite{Flauger:2008ad} also have vanishing extremal couplings \cite{Revello:2026opc}. As a possible bulk interpretation, the non-vanishing of the cubic couplings can be traced to the presence of O-planes wrapping cycles in distinct homology classes. 

Motivated by these observations, in this paper we show that in the supersymmetric DGKT scenario on a \textit{generic} Calabi--Yau orientifold, the cubic couplings for (super-)extremal configurations vanish if and only if the Calabi--Yau is rigid, i.e.~$h^{2,1}=0$. This generalises the analysis performed for toroidal orbifolds \cite{Bobev:2025yxp,Revello:2026eqp,Revello:2026opc}, and further supports the picture that the cubic couplings vanish precisely when the O-planes wrap three-cycles within a single homology class. Indeed, a generic Calabi--Yau threefold admits $h^{2,1}+1$ orientifold-even homology classes of three-cycles.

We perform our analysis in multiple steps, using new insight into the structure and interpretation of spectra and of cubic couplings for $\mathcal{N}=1$ theories. First, in section \ref{sec:real_variables}, we explain how in generic $\mathcal{N}=1$ AdS supergravity theories that are described by a real superpotential $P$ (which, in 4d, can be constructed from a K\"ahler invariant combination of the K\"ahler- and superpotential), the eigenvalues of the Hessian of this real superpotential $P$ compute the conformal dimensions of the dual operators, and we therefore refer to this matrix as the dimension matrix. Next, we show that for extremal arrangements, the cubic couplings vanish if the corresponding cubic derivative of the superpotential $P$ vanishes as well. Then, in section \ref{sec:complex_variables}, we show that the spectrum of operators can be determined rather nicely in the context of 4d $\mathcal{N}=1$ supergravity in terms of complex fields. 
With these tools, we are able to diagonalise the dimension matrix for the DGKT scenario on any Calabi-Yau three-fold in section \ref{sec:DGKT_spectrum}, consistent with, but easier than ref.~\cite{Apers:2022tfm}. Finally, in section \ref{sec:DGKT_cubic}, we show that the extremal cubic couplings always vanish for the K\"ahler sector and one universal CS/dilaton combination -- which we name the K\"ahler + universal CS/dilaton sector --  but that there are always non-vanishing (super-)extremal cubic couplings involving scalars in that sector and the remaining complex structure sector. This proves that the holographic constraint on cubic couplings of \cite{Bobev:2025yxp} is satisfied in DGKT scenarios if and only if $h^{2,1}=0$, and hence suggests that DGKT vacua arising from non-rigid Calabi-Yau manifolds belong to the swampland. Moreover, it verifies the observation of ref.~\cite{Revello:2026eqp} that the constraint is only satisfied if the O-planes wrap cycles within the same homology class, as there is only one orientifold-even homology class of three-cycles for rigid Calabi-Yau three-folds.

\section{The spectrum and cubic couplings using real fields}\label{sec:real_variables}
In $\mathcal{N}=1$ theories in three and four dimensions, the dynamics of the scalar sector is typically dictated by a superpotential. For certain higher-dimensional or supersymmetric supergravity theories the scalar sector may also admit such a description, typically in consistent truncations or BPS sectors. Moreover, an analogous formulation can arise in non-supersymmetric settings exhibiting fake supersymmetry, in which case one refers to a fake superpotential. The action of the scalar sector of such theories in $d+1$ dimensions is given by 
\begin{equation}
    S = \frac{\Mpl^{d-1}}{2}\int \dd^{d+1}x \sqrt{-g_{d+1}}\left(R_{d+1} -\frac{d-1}{2}g_{mn}(\varphi) \partial \varphi^m \partial \varphi^n - (d-1)V(\varphi) \right)\,,
\end{equation}
where $g_{mn}$ is the metric in scalar field space, parametrised by the scalar fields $\{\varphi^m\}$ and the scalar potential $V$ which can be written in terms of a real superpotential $P$:\footnote{The potential has mass dimension 2 here, and we pulled out a dimension-dependent factor such that the on-shell value of $V$ and $P$ are $-d/L_\mathrm{AdS}^2$ and $1/L_\mathrm{AdS}$ respectively.}
\begin{equation}\label{eq:V_ito_P}
    V = 2 P_m g^{mn} P_n - d\; P^2\,.
\end{equation}
Throughout the paper, we use an index notation for field derivatives, i.e.~$P_m \equiv \partial_m P$.
Note that 4d $\mathcal{N}=1$ theories admit a complex scalar field space, for which the `real superpotential' is a K\"ahler invariant function of the K\"ahler potential $K$ and holomorphic superpotential $W$,
\begin{equation}
    P = \e^\frac{K}{2} |W|\,.
\end{equation}
The scalar potential can also be written in terms of complex fields (we use $I$ ($\bar{I}$) for (anti-) \linebreak holomorphic field derivatives):
\begin{equation}
    V = 4 P_I K^{I \bar J} P_{\bar{J}} - d\; P^2\,.
\end{equation}
In the literature, one defines often $\mathcal{G} = \log P^2$. We discuss the problem in terms of complex fields in the next section. Throughout this paper, we assume that the supergravity theory allows for a vacuum solution, such that for all $\{\varphi^m\}$, the first field derivatives of the superpotential vanish:
\begin{equation}\label{eq:susy_vacuum}
    P_m = 0\,.
\end{equation}
All quantities of interest in this paper are evaluated at this supersymmetric vacuum.

\subsubsection*{The mass spectrum}
Since we have assumed the supergravity theory allows for a vacuum solution, we can calculate the mass spectrum in the said SUSY vacuum. To this aim, we should diagonalise the mass matrix, which is determined by the Hessian of the scalar potential in the frame where the kinetic terms are canonical. In the supersymmetric vacuum, (i.e.~after enforcing eq.~\eqref{eq:susy_vacuum}), we obtain
\begin{equation}\label{eq:Hessian_V}
    V_{kl} = 4 P_{km} g^{mn}P_{nl} - 2d P_0 P_{kl}\,,
\end{equation}
where $P_0$ is the value of the superpotential at the vacuum. Importantly, we assume that the spacetime is AdS and therefore take $P_0 \neq 0$. Moreover, $P_0 = L_\mathrm{AdS}^{-1}$ is the inverse AdS radius. To proceed, we write the field space metric in the vacuum in terms of vielbeins as
\begin{equation}
    g_{mn} = e^a_m e^b_n \delta_{ab}\,,
\end{equation}
and we denote their inverses as $e^m_a$. These allow us to transform the scalars to the canonical scalar fields in the vacuum. Using the following redefinitions:
\begin{equation}\label{eq:definition_Msq_and_Delta_real}
    \mathcal{M}^2_{ab} = e^k_a e^l_b V_{kl}\,, \qquad {\bf \Delta}_{ab} = e^k_a e^l_b \frac{2}{P_0} P_{kl}\,,
\end{equation}
we notice that equation \eqref{eq:Hessian_V} can be written as 
\begin{equation}
    \mathcal{M}^2 L_\mathrm{AdS}^2 = {\bf \Delta} \cdot ({\bf \Delta} - d\; \mathds{1})\,.
\end{equation}
Of course, this is very suggestive notation. The basis of eigenvectors $\{y^a_\mathbf{k}\}$ that diagonalise the matrix ${\bf \Delta}$, also diagonalise the mass matrix $\mathcal{M}^2$, indicating that the eigenvalues $\{\lambda_\mathbf{k}\}$ of the matrix ${\bf \Delta}$ are either the conformal dimension $\Delta_{\mathbf{k}}$ of the putative dual operator, or $d- \Delta_{\mathbf{k}}$:
\begin{equation}
    \lambda_{\mathbf{k}} = \Delta_{\mathbf{k}} \qquad \mathrm{or} \qquad  \lambda_{\mathbf{k}} = d - \Delta_{\mathbf{k}}\,.
\end{equation}
From here onwards, we use bold-face indices to indicate the quantities in the mass basis.

To our knowledge, this simple observation has not been made in the literature before, and it lies at the basis of our analysis of cubic couplings in DGKT in the later subsections. 

We also want to remark that the eigenvalue problem for $\mathbf{\Delta}$ can be expressed in terms of the non-canonical fields as well. When we define ${\bf \Delta}_{kl} = 2P_{kl}/P_0$ without the vielbeins, we can contract the eigenvalue equation with the inverse vielbeins such that
\begin{equation}\label{eq:eigenproblem_with_g}
    ({\bf \Delta}_{ab} - \lambda_\mathbf{i} \delta_{ab})y^b_\mathbf{i} =0 \iff ({\bf \Delta}_{kl} - \lambda_\mathbf{i} g_{kl})y^k_\mathbf{i} = 0\,,
\end{equation}
where $y_\mathbf{i}^b = y_\mathbf{i}^k e^b_k$. So in terms of the ordinary field derivatives, the eigenvalue problem is expressed in terms of the field metric instead of the identity. We put this to use later.

\subsubsection*{Cubic couplings}
The quantities that we are interested in are the cubic couplings that determine the three-point functions of bulk scalar operators, as they are the ones that are relevant for the constraint of ref.~\cite{Bobev:2025yxp}. 
These can be obtained by expressing the action in terms of the mass diagonalised fields, which we denote with bold-face indices $\mathbf{i},\mathbf{j},...$ as $\{\varphi^\mathbf{i}\}$.
When expanding the action around the vacuum, $\varphi^\mathbf{i} = \tilde \varphi^\mathbf{i} + \varphi^\mathbf{i}_0$, and truncating from quartic order onwards, we find
\begin{multline}
    S = \frac{\Mpl^{d-1}}{2}\int \dd^{d+1}x \sqrt{-g}\biggl(R - (d-1)\Bigl[\frac{1}{2}\delta_{\mathbf{ij}}\partial \tilde\varphi^\mathbf{i}\partial\tilde\varphi^\mathbf{j}+ \frac{d_\mathbf{klm}}{2} \tilde\varphi^\mathbf{k}\partial \tilde\varphi^\mathbf{l}\partial\tilde\varphi^\mathbf{m}\\-\frac{d}{L_\mathrm{AdS}^2} + \frac{m_\mathbf{i}^2}{2} (\tilde\varphi^\mathbf{i})^2 + \frac{c_\mathbf{klm}}{3!}\tilde\varphi^\mathbf{k} \tilde\varphi^\mathbf{l}\tilde\varphi^\mathbf{m} + \cdots \Bigr]\biggr)\,.
\end{multline}
The coefficients $c_\mathbf{klm}$ and $d_\mathbf{klm}$ determine the coefficients of three-point functions of bulk operators in the following combination:
\begin{equation}\label{eq:definition_cprime}
    c'_\mathbf{klm} = c_\mathbf{klm} + \left[ \frac{m^2_\mathbf{k} - m^2_\mathbf{l} -m^2_\mathbf{m}}{2} d_\mathbf{klm} + \mathrm{cyclic}\right]\,.
\end{equation}
It has long been known that, whenever the dual CFT operators satisfy the extremality condition $\Delta_\mathbf{k}=\Delta_\mathbf{l}+\Delta_\mathbf{m}$ exactly, there is a kinematic divergence unless the cubic coupling \eqref{eq:definition_cprime} vanishes \cite{DHoker:1999jke,Aprile:2020uxk}. An analogous statement also holds in the super-extremal case, where $\Delta_\mathbf{k}=\Delta_\mathbf{l}+\Delta_\mathbf{m} +2n$, $n \in \mathbb{N}_0$. In \cite{Bobev:2025yxp}, it was shown that, under suitable assumptions, the same conclusion holds even when the (super-)extremality condition is satisfied up to corrections which are suppressed in a large central charge expansion, i.e.~when the operators acquire a non-zero anomalous dimension. This is precisely the condition we will apply and check in this work. The two couplings $c_\mathbf{klm}$ and $d_\mathbf{klm}$ are naturally computed from the scalar potential and field space metric as follows:
\begin{equation}\label{eq:definition_c_and_d}
   c_\mathbf{klm} = V_\mathbf{klm}\equiv y_\mathbf{k}^k y_\mathbf{l}^l y_\mathbf{m}^m \partial_k \partial_l \partial_m V \,, \qquad d_\mathbf{klm} =  y_\mathbf{k}^k y_\mathbf{l}^l y_\mathbf{m}^m \partial_k g_{lm} \, . 
\end{equation}
It is very important to note that $d_\mathbf{klm}$ is not fully symmetric; it is only symmetric in the last two indices, so the full cyclic sum does not automatically simplify. As opposed to earlier work \cite{Bobev:2025gzu,Bobev:2025yxp}, we have included the combinatorial factors in the action because we implicitly sum over all fields.

We proceed by evaluating the cubic derivative of the scalar potential in terms of the real superpotential $P$. In the supersymmetric vacuum at $P_m=0$, we find,
\begin{align}
    V_{klm} &= 2 \cdot 3! [P_{p(k}\partial_l g^{pq} P_{m)q} +  P_{p(k}g^{pq}P_{lm)q}] - 2d P_0 P_{klm}\\
     &=  3! \left[\frac{1}{2}P_0^2 \mathbf{\Delta}_{p(k}\partial_l g^{pq} \mathbf{\Delta}_{m)q} + 
     P_0  \mathbf{\Delta}_{p(k}g^{pq}P_{lm)q}\right] - 2d P_0 P_{klm}\,.
\end{align}
We have used eq.~\eqref{eq:definition_Msq_and_Delta_real} in the last line.
Then, we can use that $\partial_l g^{pq} = - g^{pr}(\partial_l g_{rs} )g^{sq}$, and by transforming this to the basis of orthonormal mass eigenstates with eq.~\eqref{eq:definition_c_and_d}, we get that 
\begin{equation}
    V_\mathbf{klm} =  3!\left[-\frac{1}{2} P_0^2 \mathbf{\Delta}_{\mathbf{p}(\mathbf{k}} d_\mathbf{l}^\mathbf{pq} \mathbf{\Delta}_{\mathbf{m})\mathbf{q}} + P_0 \mathbf{\Delta}_{\mathbf{p}(\mathbf{k}}\delta^\mathbf{pq}P_{\mathbf{lm})\mathbf{q}}\right] - 2d P_0 P_\mathbf{klm}\,.
\end{equation}
By construction, $\mathbf{\Delta}_\mathbf{kl} = \lambda_\mathbf{k} \delta_\mathbf{kl}$, which simplifies the expression above to:
\begin{equation}\label{eq:V_klm_in_P_final}
    V_\mathbf{klm} =  [- P_0^2 (\lambda_\mathbf{k}\lambda_\mathbf{l}d_\mathbf{m kl}+\lambda_\mathbf{l}\lambda_\mathbf{m}d_\mathbf{k lm}+ \lambda_\mathbf{k}\lambda_\mathbf{m}d_\mathbf{l km}) + 
     2 P_0 (\lambda_\mathbf{k}+ \lambda_\mathbf{l}+ \lambda_\mathbf{m}) P_\mathbf{klm}] - 2d P_0 P_\mathbf{klm}\,.
\end{equation}
Looking at the contribution coming from the derivative coupling in eq.~\eqref{eq:definition_cprime}, we see it can be written in terms of the eigenvalues of $\mathbf{\Delta}$ as follows:
\begin{align}
    \frac{m^2_\mathbf{k} - m^2_\mathbf{l} -m^2_\mathbf{m}}{2} d_\mathbf{klm} &= \frac{\lambda_\mathbf{k}(\lambda_\mathbf{k}-d) - \lambda_\mathbf{l}(\lambda_\mathbf{l}-d) -\lambda_\mathbf{m}(\lambda_\mathbf{m}-d)}{2L_\mathrm{AdS}^2} d_\mathbf{klm}\\
    &= \frac{\lambda_\mathbf{k}^2-( \lambda_\mathbf{l}^2+\lambda_\mathbf{m}^2) -d(\lambda_\mathbf{k}-\lambda_\mathbf{l}-\lambda_\mathbf{m})}{2L_\mathrm{AdS}^2} d_\mathbf{klm}\,.
\end{align}
Combining this with eq.~\eqref{eq:V_klm_in_P_final} for the final cubic coupling $c'_\mathbf{klm}$ in eq.~\eqref{eq:definition_cprime}, we find that 
\begin{equation}\label{eq:cprime_final}
    c'_\mathbf{klm} = \frac{2}{L_\mathrm{AdS}}[(\lambda_\mathbf{k}+ \lambda_\mathbf{l}+ \lambda_\mathbf{m})-d ]P_\mathbf{klm} + \left[\frac{\lambda_\mathbf{k}^2-( \lambda_\mathbf{l}+\lambda_\mathbf{m})^2 -d(\lambda_\mathbf{k}-\lambda_\mathbf{l}-\lambda_\mathbf{m})}{2L_\mathrm{AdS}^2} d_\mathbf{klm}+ \mathrm{cyclic}\right]\,.
\end{equation}
Remarkably, the derivative couplings coming from the potential, i.e.~the first term in eq.~\eqref{eq:V_klm_in_P_final}, complete the squares of $\lambda_\mathbf{l}+\lambda_\mathbf{m}$ for instance. 

We discuss now when the cubic couplings vanish.
First, we consider the scenario in which the terms in eq.~\eqref{eq:cprime_final}, proportional to the derivative coupling, do not contribute. This can occur when the derivative coupling $d_\mathbf{klm}$ and its cylic permutations vanishes identically.
Another possibility is that only one of the cyclic permutations, $d_\mathbf{klm}$, $d_\mathbf{lmk}$ or $d_\mathbf{mkl}$ is non-vanishing, let us assume without loss of generality that it is $d_\mathbf{klm}$. Then the term proportional to $d_\mathbf{klm}$ vanishes as well when
\begin{equation}\label{eq:sum_lambdas}
    \lambda_\mathbf{l}+\lambda_\mathbf{m} = \lambda_\mathbf{k}\,.
\end{equation}
Remarkably, this is precisely what happens for \textit{extremal} arrangements in either of the following cases:
\begin{alignat}{2}\label{eq:extremal_arrangement_type_1}
(\lambda_\mathbf{k},\lambda_\mathbf{l},\lambda_\mathbf{m})
  & = (\Delta_\mathbf{k}, \Delta_\mathbf{l}, \Delta_\mathbf{m})
  &\quad &\Rightarrow \quad \Delta_\mathbf{l}+\Delta_\mathbf{m} = \Delta_\mathbf{k}\,,\\ \label{eq:extremal_arrangement_type_2}
(\lambda_\mathbf{k},\lambda_\mathbf{l},\lambda_\mathbf{m})
  & = (d-\Delta_\mathbf{k}, d-\Delta_\mathbf{l}, \Delta_\mathbf{m})
  &\quad &\Rightarrow \quad \Delta_\mathbf{k} +\Delta_\mathbf{m} = \Delta_\mathbf{l}\,, \qquad \mathbf{l}\leftrightarrow \mathbf{m}\,. 
\end{alignat}
 So under the circumstances where the derivative couplings do not contribute, either due to extremal arrangements of the type \eqref{eq:extremal_arrangement_type_1}-\eqref{eq:extremal_arrangement_type_2} or by the identical vanishing of the derivative couplings, the final cubic coupling takes the form
\begin{equation}\label{eq:cprime_is_P}
    c'_\mathbf{klm} = \frac{2}{L_\mathrm{AdS}}[(\lambda_\mathbf{k}+ \lambda_\mathbf{l}+ \lambda_\mathbf{m})-d ]P_\mathbf{klm}\,.
\end{equation}
We conclude that for those arrangements, vanishing $P_\mathbf{klm}$ implies that $c'_\mathbf{klm}$ vanishes as well, and this is the key insight we put to use in section~\ref{sec:DGKT}. 

We can conclude even more when the cubic coupling satisfies eq.~\eqref{eq:cprime_is_P}. We list different possibilities according to whether the eigenvalues calculate the conformal dimensions $\Delta_\mathbf{i}$ or $d-\Delta_\mathbf{i}$:
\begin{enumerate}
\item $(\lambda_\mathbf{k},\lambda_\mathbf{l},\lambda_\mathbf{m})
  = (\Delta_\mathbf{k}, \Delta_\mathbf{l}, \Delta_\mathbf{m})$:\\
  For this case, the cubic coupling \eqref{eq:cprime_is_P} always vanishes if
  \begin{equation}
    \Delta_\mathbf{k} + \Delta_\mathbf{l} +\Delta_\mathbf{m} = d\,.
\end{equation}
The three-point function of such arrangements could also lead to pathologies \cite{Castro:2024cmf,Bobev:2025yxp}, hence the vanishing of the cubic coupling under these circumstances makes the three-point function vanish and removes a potential pathology.
Furthermore, if the arrangement would also be extremal of the type \eqref{eq:extremal_arrangement_type_1}, then it can only be so when
\begin{equation}
    \Delta_\mathbf{k} = \frac{d}{2}\,.
\end{equation}
This is also an interesting observation, as ambiguities in computing correlation functions may arise whenever one of the conformal dimensions is equal to $d/2$.
\item$(\lambda_\mathbf{k},\lambda_\mathbf{l},\lambda_\mathbf{m})= (d-\Delta_\mathbf{k}, \Delta_\mathbf{l}, \Delta_\mathbf{m})$:\\
In this case only two of the eigenvalues compute the conformal dimensions directly, and we have taken $\lambda_\mathbf{l}$ and $\lambda_\mathbf{m}$ without loss of generality.
The cubic coupling~\eqref{eq:cprime_is_P} vanishes irrespective of the vanishing of $P_\mathbf{klm}$, when
\begin{equation}
    \Delta_\mathbf{l}+\Delta_\mathbf{m} = \Delta_\mathbf{k}\,,
\end{equation}
so when the arrangement is extremal again.
\item$(\lambda_\mathbf{k},\lambda_\mathbf{l},\lambda_\mathbf{m}) = (d-\Delta_\mathbf{k}, d-\Delta_\mathbf{l}, \Delta_\mathbf{m})$:\\
In this case, only one of the eigenvalues is equal to the conformal dimension, and without loss of generality we take $\lambda_\mathbf{m}=\Delta_\mathbf{m}$. Then, the cubic coupling~\eqref{eq:cprime_is_P} always vanishes when
\begin{equation}
    d + \Delta_\mathbf{m}-(\Delta_\mathbf{k}+\Delta_\mathbf{l})=0.
\end{equation}
Similarly to the first case, the arrangement is extremal with $\Delta_\mathbf{l}=\Delta_\mathbf{k}+\Delta_\mathbf{m}$ as in \eqref{eq:extremal_arrangement_type_2}, when also
\begin{equation}
    \Delta_\mathbf{k} = \frac{d}{2}\,.
\end{equation}
\item $(\lambda_\mathbf{k},\lambda_\mathbf{l},\lambda_\mathbf{m}) = (d-\Delta_\mathbf{k}, d-\Delta_\mathbf{l}, d-\Delta_\mathbf{m})$:\\
In this case, the cubic coupling~\eqref{eq:cprime_is_P} vanishes when 
\begin{equation}
    \Delta_\mathbf{k}+\Delta_\mathbf{l}+\Delta_\mathbf{m} = 2d\,,
\end{equation}
and when it is also extremal, one of the scalars must be marginal.
\end{enumerate}

A second possibility is that only the derivative couplings $d_\mathbf{klm}$ contribute because $P_\mathbf{klm}$ is trivially vanishing. This happens when the potential has exactly flat directions. The last possibility is that both the derivative couplings $d_\mathbf{klm}$ and $P_\mathbf{klm}$ do contribute. Then, the cubic coupling can only vanish due to a non-trivial cancellation of the two terms.

Finally, we make the observation that the cubic coupling $c'_\mathbf{klm}$ is actually the third \textit{covariant} field derivative of the scalar potential when evaluated at the vacuum:
\begin{equation}
    c'_\mathbf{klm} = \nabla_\mathbf{k}\nabla_\mathbf{l}\nabla_\mathbf{m}V\,.
\end{equation}
Indeed, we have
\begin{align}
    \nabla_k \nabla_l \nabla_m V &= V_{klm}- \Gamma^r_{lm}V_{kr} - \Gamma^r_{kl}V_{mr} - \Gamma^r_{mk}V_{lr} - (\nabla_k \Gamma^r_{lm})V_r+\Gamma^r_{lm}\Gamma^s_{kr}V_s\\ 
    \label{eq:nabla^3V_Christoffel}
    &= V_{klm} - \Gamma^r_{lm}V_{kr} - \Gamma^r_{kl}V_{mr} - \Gamma^r_{mk}V_{lr}\,.
\end{align}
We dropped terms with single derivatives on $V$ as we evaluate this quantity in the vacuum. 
Let us focus on the second term:
\begin{equation}
    \Gamma^r_{lm}V_{kr} =  V_{kr} \frac{1}{2} g^{rs} \left(\partial_l g_{sm} +\partial_m g_{sl} - \partial_s g_{lm} \right)\,.
\end{equation}
We transform this to the mass basis by contracting this with $y^k_\mathbf{k}y^l_\mathbf{l}y^m_\mathbf{m}$, and by using \linebreak that $(V_{rk} - m^2_\mathbf{k} g_{rk}) y^k_\mathbf{k}=0$ as in eq.~\eqref{eq:eigenproblem_with_g}, we find
\begin{equation}\label{eq:Gamma_V_is_m_d}
(y^k_\mathbf{k}y^l_\mathbf{l}y^m_\mathbf{m})\Gamma^r_{lm}V_{kr} = \frac{m^2_\mathbf{k}}{2} \left(d_\mathbf{lmk} + d_\mathbf{mkl}-d_\mathbf{klm}\right)\,.
\end{equation}
The last two terms in eq.~\eqref{eq:nabla^3V_Christoffel} are the same as \eqref{eq:Gamma_V_is_m_d} upon the cyclic permutation of $(\mathbf{k},\mathbf{l},\mathbf{m})$. Combining everything and rearranging the terms, we find
\begin{equation}
    \nabla_\mathbf{k}\nabla_\mathbf{l}\nabla_\mathbf{m}V = c_\mathbf{klm} + \left[ \frac{m^2_\mathbf{k} - m^2_\mathbf{l} -m^2_\mathbf{m}}{2} d_\mathbf{klm} + \mathrm{cyclic}\right]\,,
\end{equation}
which is precisely the cubic coupling $c'_\mathbf{klm}$. Although we are not aware of this observation being made in the literature before, we remark that this is to be expected as the cubic coupling must transform as a tensor in field space. Moreover, one can check that the cubic coupling $c'_\mathbf{klm}$ always vanishes if $\nabla_\mathbf{k}\nabla_\mathbf{l}\nabla_\mathbf{m}P =0$ when evaluated in the supersymmetric vacuum, independently of whether the coupling is extremal or not.

\section{The spectrum using complex fields}\label{sec:complex_variables}
We rederive some aspects of the spectrum when the supergravity theory admits a formulation in terms of complex fields and a K\"ahler- and superpotential.\footnote{See \cite{Bachlechner:2012at} for a similar analysis from a bulk supergravity (rather than CFT) point of view.} We denote the complex indices by capital letters $I$, $J$, $K$ etc. The potential, in terms of the real superpotential $P = \exp(K/2) |W|$, is given by
\begin{equation}
    V = 4P_I K^{I \bar{J}} P_{\bar{J}} - d P^2\,.
\end{equation}
One can show that the second derivatives of the potential in the supersymmetric vacuum $P_I = 0$, $P_{\bar{J}} =0$, are given by
\begin{align}\label{eq:complex_hessian_potential}
\begin{split}
    V_{KL} &= \overline{V_{\bar K \bar L}} = 4 P_{KI}K^{I \bar J} P_{\bar J L} + 4 P_{L I}K^{I \bar J} P_{\bar JK} - 2d P_0 P_{KL},\\
    V_{K \bar{L}} &= \overline{V_{\bar K L}} = 4 P_{K I} K^{I \bar J} P_{\bar J \bar L} + 4 P_{\bar L I} K^{I \bar J} P_{\bar J K}- 2d P_0 P_{K \bar L}\,.
\end{split}
\end{align}
It is again useful to write the K\"ahler metric in terms of the vielbein $\kappa$ and denote flat indices with $A$, $\bar{B}$, ... as follows:
\begin{equation}
    K_{I \bar J} = \kappa_I^A \kappa_{\bar{J}}^{\bar{B}} \delta_{A \bar B}\,, \qquad \delta_{A \bar B} = \kappa^I_A \kappa^{\bar{J}}_{\bar{B}}K_{I \bar J}\,.
\end{equation}
Next, we introduce the following block matrices:
\begin{equation}
\label{eq:complex_mass_delta_matrices}
    \mathcal{M}^2 = 
    \begin{pmatrix}
        V_{A \bar{B}} &V_{AB} \\
        V_{\bar A \bar B} & V_{\bar{A} B}
    \end{pmatrix}\,,
    \qquad
    \mathbf{\Delta} = \frac{2}{P_0}\begin{pmatrix}
        P_{A \bar{B}} & P_{AB}\\
        P_{\bar A \bar B} & P_{\bar{A} B}
    \end{pmatrix}\,, 
\end{equation}
where we have used the short notation $P_{AB} = \kappa_I^A\kappa_J^B P_{IJ}$ etc. Then, multiplying equation~\eqref{eq:complex_hessian_potential} with  $P_0^{-2}=L_\mathrm{AdS}^2$, and contracting it with two inverse vielbeins, equation~\eqref{eq:complex_hessian_potential} can be written as the following matrix identity:
\begin{equation}
    \mathcal{M}^2 L_\mathrm{AdS}^2 =  \mathbf{\Delta} \cdot (\mathbf{\Delta} - d\; \mathds{1})\,.
\end{equation}
We conclude as before that the matrix $\mathbf{\Delta}$ has the same eigenvectors as  $\mathcal{M}^2 L_\mathrm{AdS}^2 $, and that the eigenvalues of the former determine the eigenvalues of the latter.
Contrary to the analysis of section~\ref{sec:real_variables}, we can say a bit more about the spectrum. This all relies on the fact that, in the supersymmetric vacuum, the mixed holomorphic and anti-holomorphic derivatives of the real superpotential satisfy
\begin{equation}\label{eq:P_KbarL}
    P_{K \bar L} =  \frac{P_0}{2} K_{K \bar L}\,, \qquad P_{A \bar B} = \frac{P_0}{2} \delta_{A \bar B}\,.
\end{equation}
With this observation, we can write the matrix $\mathbf{\Delta}$ as
\begin{equation}
    \mathbf{\Delta} =
\begin{pmatrix}
        \delta_{A\bar B} &\mathcal{P}_{AB}\\
        \mathcal{P}_{\bar A \bar B} & \delta_{\bar A B}
    \end{pmatrix}\,, \qquad
    \mathcal{P}_{AB} = \frac{2}{P_0} P_{AB}\,, \qquad \mathcal{P}_{\bar A \bar B} = \frac{2}{P_0} P_{\bar A \bar B} = \overline{\mathcal{P}_{AB}}\,,
\end{equation}
where the latter holds because $P$ is real. The eigenvalues $\{\lambda_{\mathbf{I}\pm}\}$ and eigenvectors $\{y_{\mathbf{I}\pm}\}$ of the dimension matrix $\mathbf{\Delta}$ can then be expressed in terms of those of $\mathcal{P}$. When $\mathcal{P}$ is symmetric and real, which is the relevant case in the next sections, its eigenvalues $\{\lambda_\mathbf{I}\}$ are real, and the eigenvalues of $\overline{\mathcal{P}}$ are the same.\footnote{In the case that $\mathcal{P}$ is not real (but still symmetric), we can use the singular value decomposition of $\mathcal{P}$. The eigenvalues $\lambda_{\mathbf{I}_\pm}$ are then equal to $1 \pm \sigma_\mathbf{I}$, where the $\{\sigma_\mathbf{I}\}$ are the singular values of $\mathcal{P}$. The full eigenvectors can be written in terms of the singular vectors of $\mathcal{P}$, i.e. $y_{\mathbf{I}_\pm} = \left(\pm \bar{x}_\mathbf{I}^{\bar{B}}, y_\mathbf{I}^B\right)$ with $\mathcal{P}_{AB} y_\mathbf{I}^B = \sigma_\mathbf{I} \delta_{A \bar B}\bar{x}^{\bar{B}}_\mathbf{I}$ and $\bar{\mathcal{P}}_{\bar{A} \bar{B}} \bar{x}_\mathbf{I}^{\bar B} = \sigma_\mathbf{I} \delta_{\bar A  B} y^{B}_\mathbf{I}$.} We then distinguish between the ``holomorphic'' and ``antiholomorphic'' eigenvectors, namely $\{y^B_\mathbf{I}\}$ for $\mathcal{P}_{AB}$ and $\{\bar{y}^{\bar B}_\mathbf{I}\}$ for $\mathcal{P}_{\bar A \bar B}$. The eigenvalues and eigenvectors of $\mathbf{\Delta}$ can then be written as
\begin{equation}\label{eq:eigensystem_complex}
    \lambda_{\mathbf{I}_\pm} = 1 \pm \lambda_\mathbf{I}\,, \qquad y_{\mathbf{I}_\pm} =  \eta^\pm \left(\pm \bar{y}_\mathbf{I}^{\bar{B}}, y_\mathbf{I}^B\right)\,.
\end{equation}
We have introduced $\eta^+ =1$, $\eta^- = i$ such that the eigenvectors \{$y_{\mathbf{I}_+}$\}  and \{$y_{\mathbf{I}_-}$\} correspond to the saxionic and axionic eigenvectors, respectively. Accordingly, the index sets $\{\mathbf{I}_+\}$ and $\{\mathbf{I_-}\}$ label the real scalar degrees of freedom and can be identified with the mass eigenstate indices $\{\mathbf{i}\}$ introduced in the previous section such that $\{\mathbf{i}\} = \{\mathbf{I}_{+}, \mathbf{I}_{-}\}$. Hence, we note that contractions over real variables can be expressed in complex variables as follows:
\begin{equation}\label{eq:fromRealtoComplexContraction}
    X_{...b}y^b_{\mathbf{I}_\pm} = \eta^\pm\left(X_{...B}y^B_{\mathbf{I}}\pm X_{...\bar B}\bar{y}^{\bar B}_{\mathbf{I}}\right)\,.
\end{equation}
As before, the eigenvalues determine the conformal dimensions, by either
\begin{equation}
    \lambda_{\mathbf{I}_\pm} = \Delta_{\mathbf{I}_\pm}\,, \quad \mathrm{or} \quad \lambda_{\mathbf{I}_\pm} = d - \Delta_{\mathbf{I}_\pm}\,.
\end{equation}
However, an important observation is that
\begin{equation}
    \lambda_{\mathbf{I}_-} + \lambda_{\mathbf{I}_+} =2\,.
\end{equation}
From the unitarity bound $\Delta \geq (d-2)/2$, it is clear that this means that $\lambda_{\mathbf{I}-}$ and $\lambda_{\mathbf{I}_+}$ can only be both the conformal dimensions when $|\lambda_\mathbf{I}|\leq (4-d)/2$. 
When $|\lambda_\mathbf{I}|> (4-d)/2$, either $\lambda_{\mathbf{I}_+}$ or $\lambda_{\mathbf{I}_-}$ must be $\Delta$, and the other is $d-\Delta$. In that case, we see that 
\begin{equation}
    |\Delta_{\mathbf{I}_+}-\Delta_{\mathbf{I}_-}| = |\lambda_{\mathbf{I}_+}+\lambda_{\mathbf{I}_-}-d|= |d-2|.
\end{equation}

We end this section with two remarks. First, we comment on the eigenvalue problem for $\mathcal{P}$. As mentioned earlier, this consists of finding the eigenvectors $\{v_\mathbf{I}^B\}$ and eigenvalues $\{\lambda_\mathbf{I}\}$ such that
\begin{equation}
    \left(\mathcal{P}_{AB} - \lambda_\mathbf{I}\delta_{AB}\right)y^B_\mathbf{I} = 0\,.
\end{equation}
Multiplying with the inverse vielbeins of the K\"ahler metric, we find 
\begin{equation}
\label{eq:eigenvalue_problem_K}
    \left(\mathcal{P}_{IJ} - \lambda_\mathbf{I} \tilde K_{IJ}\right)y^J_\mathbf{I} = 0\,, \qquad \tilde K_{IJ} = \delta_{AB}\kappa_I^A \kappa^J_B\,, \qquad y^J_\mathbf{I} = y^B_\mathbf{I} \kappa_B^J\,. 
\end{equation}
In the parametrisation where saxions are the real part of the complex fields, all holomorphic and anti-holomorphic derivatives of the K\"ahler potential are the same, and $\tilde{K}_{IJ} = K_{IJ} = K_{I \bar J}$.
Solving this eigenvalue problem turns out to be easier, as one does not require to know the vielbeins of the K\"ahler metric, which can become rather complicated.  We put this to use in our analysis for the spectrum and cubic couplings of DGKT in the next section.

Second, we note that the analogue of eq.~\eqref{eq:P_KbarL} involving three derivatives also holds:
\begin{equation}\label{eq:P_KbarLbarM}
    P_{KL\bar M} = \frac{P_0}{2} K_{KL\bar M}\,, \qquad P_{K \bar L\bar M} = \frac{P_0}{2} K_{K\bar L\bar M}\,.
\end{equation}
We also use this extensively in the next section.

\section{DGKT models, their spectrum and cubic couplings}\label{sec:DGKT}
We apply the insights of section~\ref{sec:real_variables} and~\ref{sec:complex_variables} for the DGKT scenario \cite{DeWolfe:2005uu}, which is an AdS$_4$ compactification of massive type IIA string theory. For the compactification manifold, we consider a \textit{general} Calabi-Yau three-fold with $h^{1,1}_-$ K\"ahler moduli $\{T^\alpha\}$ with general triple intersection numbers $\{\kappa_{\alpha\beta\gamma}\}$, $h^{2,1}$ complex structure moduli and the axion-dilaton which we collectively denote by $\{U^\mu\}$, and we denote indices now by $I= (\alpha, \mu)$ etc. We work with the conventions that the real (imaginary) parts of these scalars are the saxionic (axionic) degrees of freedom.
The full K\"ahler potential is the sum of the K\"ahler and complex structure ones, $K = K_\mathrm{KM} + K_\mathrm{Q}$, where the latter also contains the dilaton.
The K\"ahler potential $K_\mathrm{KM}$ is determined entirely in terms of the triple intersection numbers $\kappa_{\alpha\beta\gamma}$ as follows
\begin{equation}
    K_\mathrm{KM} = - \log \left[\kappa_{\alpha\beta\gamma} (T^\alpha + \bar T^\alpha)(T^\beta + \bar T^\beta)(T^\gamma + \bar T^\gamma)\right]\,.
\end{equation}
 while the K\"ahler potential $K_\mathrm{Q}$ can be written as
 \begin{equation}
    K_\mathrm{Q}=-\frac{4}{n} \log \left[ P_n \left( \frac{U^{\mu}+\bar{U}^{\mu}}{2}\right)\right] \qquad \qquad \mu=1\cdots n,
\end{equation}
where $n=h^{2,1}+1$ and $P_n(z^{\mu})$ is a positive, homogeneous function of degree $n$. Alternatively, one may trade one such degree of freedom for the four-dimensional dilaton $D$ through
\begin{equation}
 P_n \left[ \frac{\e^D}{2} \left(U^{\mu}+\bar{U}^{\mu} \right) \right]= 1,
\end{equation}
which also allows us to rewrite $K_\mathrm{Q}=4D$. This degree of freedom is always present, also when $h^{2,1}=0$.
Importantly, the full K\"ahler potential only depends on the real part of the complex moduli, and it satisfies the following no-scale relations:
\begin{align}\label{eq:no_scale_relations}
\begin{split}
    &K_\alpha (T^\alpha + \bar T^\alpha) = - 3\,, \qquad \quad \; K_\mu (U^\mu+ \bar U^\mu) = -4 \,,\\
    &K_{\alpha \beta}(T^\beta + \bar T^\beta) = -K_\alpha\,, \qquad K_{\mu \nu} (U^\nu+ \bar U^\nu) = - K_\mu\,,\\
    &K_{\alpha \beta \gamma}(T^\gamma + \bar T^\gamma) = -2K_{\alpha\beta}\,, \;\;\, K_{\mu \nu \rho} (U^\rho+ \bar U^\rho) = - 2K_{\mu\nu}\,,\\
    &\kappa_{\alpha \beta \gamma} (T^\beta + \bar T^\beta)(T^\gamma + \bar T^\gamma) = - \frac{1}{3}\e^{-K_\mathrm{KM}} K_\alpha\\
    &\kappa_{\alpha \beta \gamma} (T^\gamma + \bar T^\gamma) = -\frac{1}{6}\e^{-K_\mathrm{KM}}\left( K_{\alpha\beta}-K_\alpha K_\beta\right)\,, \\
\end{split}
\end{align}
The second and third lines follow from taking derivatives of the first line.
We parametrise the DGKT superpotential $W$ by
\begin{equation}\label{eq:superpotential_DGKT}
    W = e_\alpha T^\alpha + m_0 \kappa_{\alpha\beta\gamma}T^\alpha T^\beta T^\gamma - 2 p_\mu U^\mu\,,
\end{equation}
where all the coefficients $\{e_\alpha, m_0, p_\mu\}$ are real and refer to the $F_4$-, $F_0$- and $H$-flux parameters respectively. In the literature, as in the original DGKT paper \cite{DeWolfe:2005uu}, sometimes also $F_2$- and $F_6$-flux parameters are included, inducing vevs for the axions but chosen such that the physical $F_2$ and $F_6$ fluxes vanish. These setups are related to the superpotential above by a field redefinition, namely imaginary shifts, corresponding to $\mathrm{SL}(2,\mathbb{C})$ transformations of the moduli-space coordinates.\footnote{A more general form of the superpotential would include the terms $i e^0$ and $3 i \kappa_{\alpha\beta\gamma} m^\alpha T^\beta T^\gamma$, where $\{m^\alpha\}$ and $e_0$ parametrise the Page two- and six-form fluxes. These can be obtained by the field redefinitions \linebreak $T^\alpha \to T^\alpha + i m^\alpha/m_0$ and $U^\mu \to  U^\mu + i c^\mu$, combined with a relabelling of the other flux parameters, i.e. $ e_\alpha^\mathrm{new} = e_\alpha - 3 \kappa_{\alpha \beta \gamma}m^\beta m^\gamma / m_0$ and $\hat e_0^\mathrm{new} = -2 p_\mu c^\mu + e_\alpha m^\alpha/m_0 -\kappa_{\alpha \beta \gamma}m^\alpha m^\beta m^\gamma/m_0^2 $.} Similarly, for the toroidal orbifolds considered in refs.~\cite{Banks:2006hg,Cribiori:2021djm,Carrasco:2023hta}, a double T-duality can be represented as a $\mathrm{SL}(2,\mathbb{Z})$ transformation acting on the relevant Kähler modulus. The associated Jacobian is diagonal, hence our final results are independent of such coordinate transformations, and it allows us to work in the current frame with the superpotential above.

From the F-term equations $W_I + K_I W_0=0$ and the no-scale relations \eqref{eq:no_scale_relations}, we note that
\begin{align}\label{eq:F-term_no-scale_U}
    &(U^\mu+ \bar U^\mu)\partial_\mu W = 4 W_0\\
    \label{eq:F-term_no-scale_T}
    &(T^\alpha + \bar T^\alpha) \partial_\alpha W = 3 W_0\,,
\end{align}
and the same holds for the complex conjugates. Equation \eqref{eq:F-term_no-scale_U} and its conjugate imply that $W_0 = \overline{W}_0$ is real, as the $\{p_\mu\}$ are real. Following ref.~\cite{DeWolfe:2005uu}, the F-term equation for the K\"ahler moduli implies that 
\begin{equation}
    e_\alpha + 3 m_0 \kappa_{\alpha \beta \gamma} T^\beta T^\gamma + W_0 K_\alpha = 0.
\end{equation}
Since $e_\alpha$, $m_0$, $\kappa_{\alpha\beta\gamma}$, $W_0$ and $K_\alpha$ are real, we have that $\kappa_{\alpha \beta \gamma} \Re(T^\beta)\Im(T^\gamma) $ must vanish. Because $\kappa_{\alpha \beta \gamma} \Re(T^\beta)$ defines a non-degenerate matrix, we must have that all $\Im(T^\gamma)$ vanish and hence that all the K\"ahler moduli vevs are real. From eq.~\eqref{eq:superpotential_DGKT}, it follows then that also $p_\mu U^\mu$ must be real, and we find that eq.~\eqref{eq:F-term_no-scale_U} implies that
\begin{equation}\label{eq:min_pU_is_W}
    -2 p_\mu U^\mu = 2 W_0\,.
\end{equation}
Combining this with eq.~\eqref{eq:superpotential_DGKT}, and with \eqref{eq:F-term_no-scale_T}, we find for the SUSY vacuum that
\begin{align}
     &e_\alpha T^\alpha + m_0 \kappa_{\alpha\beta\gamma}T^\alpha T^\beta T^\gamma = -W_0\\
   & e_\alpha (T^\alpha + \bar T^\alpha) + 3 m_0 \kappa_{\alpha\beta\gamma}(T^\alpha +\bar T^\alpha)T^\beta T^\gamma = 3 W_0\,.
\end{align}
Having established that the K\"ahler moduli are real, i.e.~$T^\alpha = \bar T^\alpha$ we find that the on-shell values of the other terms in the superpotential are:
\begin{equation}\label{eq:on-shell_terms_W}
    e_\alpha T^\alpha = - \frac{9}{4} W_0\,, \qquad m_0 \kappa_{\alpha\beta\gamma}T^\alpha T^\beta T^\gamma = \frac{m_0}{8}\e^{-K_\mathrm{KM}}= \frac{5}{4} W_0\,.
\end{equation}

\subsection{The spectrum}\label{sec:DGKT_spectrum}
We are now ready to calculate the scalar spectrum. Note that this has already been done in ref.~\cite{Apers:2022tfm}, but we show that the new techniques outlined in section~\ref{sec:complex_variables} provide a more efficient method.

As argued before, we need to solve eq.~\eqref{eq:eigenvalue_problem_K}. We note first that, in terms of the K\"ahler and superpotential, the matrix $\mathcal{P}$ becomes:\footnote{\label{fn:P_IJ}This identity is not specific to DGKT and is valid for any supergravity theory where the real superpotential can be expressed in terms of a K\"ahler and superpotential, also when there is no notion of holomorphicity as in the 3d $\mathcal{N}=1$ formalism of ref.~\cite{VanHemelryck:2022ynr}.}
\begin{equation}
    \mathcal{P}_{IJ} = \frac{2}{P_0}P_{IJ} = \left( \frac{W_{IJ}}{W_0} + K_{IJ} - K_I K_J\right)\,,
\end{equation}
which can be easily derived from the fact that $\partial_{IJ} P^2 = 2P_0 P_{IJ}$, the definition of $P$ and the F-term equations.
The superpotential is linear in the fields $U^\mu$, and the only surviving second derivative is
\begin{equation}
    \frac{W_{\alpha \beta}}{W_0} = \frac{6 m_0}{W_0}  \kappa_{\alpha\beta\gamma} T^\gamma = 60 \e^{K_\mathrm{KM}} \kappa_{\alpha\beta\gamma} T^\gamma = -5\left(K_{\alpha\beta}- K_\alpha K_\beta\right)\,,
\end{equation}
where we have used eq.~\eqref{eq:on-shell_terms_W}, the fact that the K\"ahler moduli are real on-shell and the no-scale relations~\eqref{eq:no_scale_relations}.
With this, and the fact that $K_{\alpha\mu}=0$, we can write $\mathcal{P}$ in a block form:
\begin{equation}\label{eq:Hessian_P_DGKT}
    \mathcal{P} = \begin{pmatrix}
        -4(K_{\alpha\beta}-K_\alpha K_\beta) & -K_\nu K_\alpha \\
        -K_\beta K_\mu & K_{\mu \nu} - K_\mu K_\nu
    \end{pmatrix},
\end{equation}
What is remarkable, is that after these manipulations, the matrix $\mathcal{P}$ is only expressed in terms of first and second derivatives of the K\"ahler potential, all the dependence of the superpotential has been substituted away. As a consequence, the spectrum can easily be obtained using the no-scale identities \eqref{eq:no_scale_relations}.

First, we show that there exist two eigenvectors of the form
\begin{equation}
    v^J_\mathbf{1} = (T^\beta+\bar T^\beta, \xi_v (U^\nu+ \bar U^\nu))\,, \quad w^J_\mathbf{1} = (T^\beta+\bar T^\beta, \xi_w (U^\nu+ \bar U^\nu))\,,
\end{equation}
with eigenvalues $\lambda =9$ and $\lambda =-4$ corresponding to $\xi_v=-1/4$ and $\xi_w=3$ respectively.
Indeed, we see that for a vector of this type eq.~\eqref{eq:eigenvalue_problem_K} becomes
\begin{equation}
    \begin{pmatrix}
    -4\left(K_{\alpha\beta}-K_\alpha K_\beta\right)(T^\beta+\bar T^\beta) -K_\nu K_\alpha \xi (U^\nu+ \bar U^\nu) - \lambda K_{\alpha \beta } (T^\beta+\bar T^\beta)\\
    -K_{\mu}K_\beta (T^\beta+\bar T^\beta) + \xi\left(K_{\mu \nu} - K_\mu K_\nu\right)(U^\nu +\bar U^\nu) - \lambda \xi K_{\mu \nu}(U^\nu +\bar U^\nu)
    \end{pmatrix}=0\,.
\end{equation}
With the no-scale relations of eq.~\eqref{eq:no_scale_relations}, and the fact that the vevs of the fields are real, this reduces to 
\begin{equation}
    \begin{pmatrix}
    \left[-4(-1+3) + 4\xi + \lambda \right]K_{\alpha}\\
    \left[3 +\xi (-1 +4+\lambda)\right] K_\mu
    \end{pmatrix}=0\,,
\end{equation}
which vanishes indeed when $(\lambda, \xi) = (9,-1/4)$ and $(-4,3)$. 
Note that all eigenvectors corresponding to different eigenvalues must be orthogonal to each other with the K\"ahler metric defining the inner product, so 
\begin{equation}
    y_\mathbf{I} \cdot y_\mathbf{J}\equiv y^I_\mathbf{I} K_{IJ} y^J_\mathbf{J}=0  \qquad \mathrm{if} \qquad \lambda_\mathbf{I} \neq \lambda_\mathbf{J}\,.
\end{equation}
This is indeed the case for the two eigenvectors we found. One could further normalise these eigenvectors with this inner product, but we choose not to do so as it is not necessary for our purposes. Contracting them with the first derivatives of the K\"ahler potential, we find
\begin{equation}\label{eq:Kv_and_Kw}
    K_{J} v^J_\mathbf{1} = - 2\,, \qquad  K_{J} w^J_\mathbf{1} = -15\,.
\end{equation}
Next, using the no-scale relations, it is easy to see that there are $h^{1,1}_- -1$ more eigenvectors of the matrix $\mathcal{P}$ in eq.~\eqref{eq:Hessian_P_DGKT} with eigenvalue $-4$. Because $K$ is a real function of the real parts of the scalars, we can consider a basis where these eigenvectors are real and orthogonal to both of the eigenvectors above in the following way:
\begin{equation}
    w^J_\mathbf{J} = (w^\beta_\mathbf{J}, 0)\,, \qquad w^{\beta}_\mathbf{J}K_\beta = 0\,, \qquad \mathbf{J} = 2, ..., h^{1,1}_-\,.
\end{equation}
Very similarly, we can take a real basis of $h^{2,1}$ eigenvectors with eigenvalue $\lambda=1$ satisfying
\begin{equation}
    u^J_\mathbf{J} = (0, u^\nu_\mathbf{J})\,, \qquad u^{\nu}_\mathbf{J} K_\nu = 0\,, \qquad \mathbf{J} = 1,...,h^{2,1}\,.
\end{equation}
We summarise what we have obtained: we determined that the matrix $\mathcal{P}$ has one eigenvalue equal to 9, $h^{1,1}_-$ eigenvalues equal to $-4$, and $h^{2,1}$ eigenvalues equal to 1. 
As we have seen in the previous section, this means that the eigenvalues of $\mathbf{\Delta}$ come in pairs, and they in turn determine the conformal dimensions of the scalar operators. We summarise this in table~\ref{tab:spectrum} below.
\begin{table}[ht]
    \centering
    \begin{tabular}{|c|c|c|c|c|}\hline
    $\lambda_\mathbf{K}$ & $\lambda_{\mathbf{K}_-},\lambda_{\mathbf{K}_+}$ & $\Delta_{\mathbf{K}_-}$, $\Delta_{\mathbf{K}_+}$ & $y_\mathbf{K}$ & Multiplicity\\ \hline \hline
    $9$ & $-8,10$ & $11,10$ & $v_\mathbf{1}$& $1$\\ \hline
    $-4$ & $5,-3$ & $5,6$ & $w_\mathbf{K}$ & $h^{1,1}_-$\\ \hline
    $1$ & $0,2$ & $3,2$ & $u_\mathbf{K}$& $h^{2,1}$\\ \hline
    \end{tabular}
    \caption{Summary of the spectrum. The eigenvalues of $\mathcal{P}$ are denoted by $\lambda_\mathbf{K}$, those of $\mathbf{\Delta}$ by $\lambda_{\mathbf{K}_\pm}$, and the corresponding conformal dimensions by $\Delta_{\mathbf{K}_\pm}$. The eigenvectors of $\mathcal{P}$ are denoted by $y_\mathbf{K}$.}
    \label{tab:spectrum}
\end{table}\\
We refer to the scalar sector determined by $\{v_\mathbf{1}, w_\mathbf{j}\}$ as the ``K\"ahler + universal CS/dilaton sector'', as this involves all K\"ahler moduli and only one universal combination of the complex structure moduli and dilaton.
Finally, we note that we can write the eigenvectors of the dimension matrix $\mathbf{\Delta}$ using eq.~\eqref{eq:eigensystem_complex} as follows:
\begin{equation}
    v_{\mathbf{1}_\pm} = \eta_\pm \left(\pm \bar{v}_{\mathbf{1}}^{\bar{I}}, v_{\mathbf{1}}^{I}\right)\,, \qquad w_{\mathbf{I}_\pm} = \eta_\pm \left(\pm \bar{w}_{\mathbf{I}}^{\bar{I}}, w_{\mathbf{I}}^I\right)\,, \qquad u_{\mathbf{I}_\pm} = \eta_\pm\left(\pm \bar{u}_{\mathbf{I}}^{\bar{I}}, u_{\mathbf{I}}^I\right)\,.
\end{equation}
The vectors $v_{\mathbf{1}}$ and $w_{\mathbf{1}}$ are real as can be seen from their definition, and we have also chosen a real set of eigenvectors $\{w_\mathbf{J}\}$ and $\{u_\mathbf{J}\}$; i.e.~all eigenvectors are real:
\begin{equation}\label{eq:eigenvectors_are_real}
    \bar{y}_{\mathbf{I}} = y_{\mathbf{I}} \,.
\end{equation}

\subsection{Cubic couplings}\label{sec:DGKT_cubic}
In this subsection, we compute the cubic couplings for the (super-)extremal arrangements of DGKT. We start with the extremal arrangements $(10,5,5)$ and $(6,5,11)$, with scalars belonging to the K\"ahler + universal CS/dilaton sector. Looking at the spectrum summarised in table~\ref{tab:spectrum}, we see that the corresponding eigenvalues $\{\lambda_{\mathbf{K}_\pm}\}$ are such that eq.~\eqref{eq:sum_lambdas} is satisfied for these arrangements, as $5+5=10$ and $ 5 - 8=-3$. 
As we have explained in section~\ref{sec:real_variables}, the only way then for the cubic coupling to vanish, is if the third field derivative of the real superpotential, expressed in mass eigenstates, vanishes.

For extremal arrangements involving two axions and one saxion, which is the case for these arrangements, we are interested in computing the following quantity:
\begin{equation}\label{eq:P_aas_to_compute}
    P_{ijk} y^{i}_\mathbf{I_-}y^{j}_\mathbf{J_-}y^{k}_\mathbf{K_+}\,.
\end{equation}
Using eq.~\eqref{eq:fromRealtoComplexContraction} to transform this expression in terms of the complex fields, we obtain eight different terms, but they are related to each other. We illustrate this with two examples:
\begin{align}
    P_{IJ\bar K}y^{I}_\mathbf{I}y^{J}_\mathbf{J}\bar{y}^{\bar  K}_\mathbf{K} &= \frac{P_0}{2} K_{IJ\bar K}y^{I}_\mathbf{I}y^{J}_\mathbf{J}\bar{y}^{\bar  K}_\mathbf{K} = \frac{P_0}{2} K_{IJK} y^{I}_\mathbf{I}y^{J}_\mathbf{J}y^{K}_\mathbf{K}\,,\\
    P_{\bar I \bar J \bar K} \bar{y}^{\bar I}_\mathbf{I}\bar{y}^{\bar J}_\mathbf{J}\bar{y}^{\bar K}_\mathbf{K} &=P_{IJK} y^{I}_\mathbf{I}y^{J}_\mathbf{J}y^{  K}_\mathbf{K}\,.
\end{align}
We have used eq.~\eqref{eq:P_KbarLbarM} for the mixed holomorphic and anti-holomorphic derivatives of $P$ and eq.~\eqref{eq:eigenvectors_are_real}. Then, eq.~\eqref{eq:P_aas_to_compute} becomes
\begin{equation}\label{eq:P_ijk_aas}
    P_{ijk} y^{i}_\mathbf{I_-}y^{j}_\mathbf{J_-}y^{k}_\mathbf{K_+} = -\left(2P_{IJK} -P_0 K_{IJK}\right)y^{I}_\mathbf{I}y^{J}_\mathbf{J}y^{K}_\mathbf{K}\,.
\end{equation}
The final result is symmetric in $I, J, K$ and therefore depends only on the multiplet structure of the scalars, rather than on their specific arrangement within a given extremal configuration. Two extremal configurations thus yield the same result whenever their scalar operators can be paired one-to-one such that each pair belongs to the same multiplet.
For instance, $(10,5,5)$ and $(6,5,11)$ are equivalent in this sense, since their operators can be matched into the multiplets $(10,11)$ and two copies of $(5,6)$.
The relation between cubic couplings for operators belonging to the same multiplets has also been discussed in \cite{Revello:2026opc} (see also \cite{Rong:2018okz}).

We show now that this vanishes for these extremal arrangements, for which we need to evaluate $P_{IJK}w^I_\mathbf{I}w^J_\mathbf{J}v^K_\mathbf{1}$.

In the supersymmetric vacuum, we find that $\partial_{IJK} P^2 = 2 P_0 P_{IJK}$, and using the superpotential $W$ and K\"ahler potential, we find:
\begin{align}\label{eq:P_IJK}
\begin{split}
    P_{IJK} =& \;\frac{P_0}{2} \biggl(\frac{W_{IJK}}{W_0} +K_{IJK} + K_I K_J K_K -(K_{IJ}K_K + K_I K_{JK} + K_J K_{IK})\\
    &+(\mathbf{\Delta}_{IJ}K_K + \mathbf{\Delta}_{JK}K_I + \mathbf{\Delta}_{IK} K_J)\biggr)\,.
\end{split}
\end{align}
Let us focus on each term separately. The only surviving term of $W_{IJK}$ is the following:
\begin{equation}
    W_{\alpha\beta\gamma} = 6 m_0 \;  \kappa_{\alpha\beta\gamma} = 60 W_0 e^{K_\mathrm{KM}} \;  \kappa_{\alpha\beta\gamma}\,,
\end{equation}
and so we find with eq.~\eqref{eq:no_scale_relations}, we can write:
\begin{equation}
    W_{\alpha\beta K }v^K_\mathbf{1} = -10 W_0 (K_{\alpha \beta} - K_\alpha K_\beta)\,.
\end{equation}
Contracting with the other two vectors gives
\begin{equation}
    W_{IJK}(w^I_\mathbf{I}w^J_\mathbf{J}v^K_\mathbf{1}) = -10 W_0 [(w_\mathbf{I} \cdot w_\mathbf{J}) - w^\mu_\mathbf{I}K_{\mu \nu}w^\nu_\mathbf{J} -9\delta_\mathbf{I1}\delta_\mathbf{J1} ]\,,
\end{equation}
where we have used that $K_\alpha w_\mathbf{I}^\alpha = -3 \delta_\mathbf{I1}$.
For the term in the middle we use $w^\mu_\mathbf{I} = 3(U^\mu + \bar U^\mu)\delta_\mathbf{I1}$ and the no-scale relations to find
\begin{equation}
     -w^\mu_\mathbf{I}K_{\mu \nu}w^\nu_\mathbf{J} = 3\delta_\mathbf{J1} K_\mu w^\mu_\mathbf{I} = -36 \delta_\mathbf{I1}\delta_\mathbf{J1}\,.
\end{equation}
Everything combined, we find
\begin{equation}
    \frac{W_{IJK}}{W_0}(w^I_\mathbf{I}w^J_\mathbf{J}v^K_\mathbf{1}) = -10(w_\mathbf{I} \cdot w_\mathbf{J}) +450\;\delta_\mathbf{I1}\delta_\mathbf{J1}\,.
\end{equation}
For the other terms, we use eq.~\eqref{eq:Kv_and_Kw} and the orthogonality of $v_\mathbf{1}$ and $w_\mathbf{I}$ to find
\begin{alignat}{3}
    &(K_I K_J K_K) (w^I_\mathbf{I}w^J_\mathbf{J}v^K_\mathbf{1}) = -2 K_I K_J w^I_\mathbf{I}w^J_\mathbf{J} &=& -450\delta_\mathbf{I1}\delta_{\mathbf{J1}}\\
    &-(K_{IJ}K_K + K_I K_{JK} + K_J K_{IK})(w^I_\mathbf{I}w^J_\mathbf{J}v^K_\mathbf{1})\; &=& \;2 (w_\mathbf{I} \cdot w_\mathbf{J})\\
    &(\mathbf{\Delta}_{IJ}K_K + \mathbf{\Delta}_{JK}K_I + \mathbf{\Delta}_{IK} K_J) (w^I_\mathbf{I}w^J_\mathbf{J}v^K_\mathbf{1}) &=& \;8(w_\mathbf{I} \cdot w_\mathbf{J})\,.
\end{alignat}
In the last two equations, only the terms involving $K_I$ contribute as all the others vanish by the orthogonality of $v_\mathbf{1}$ with $w_\mathbf{I}$. We have also used that $w_\mathbf{J}$ is an eigenvector of (the reduced) $\Delta$ with eigenvalue $-4$ in the last equation.
Combining everything, we see that everything magically cancels, except for one term:
\begin{equation}    
     P_{IJK}(w^I_\mathbf{I}w^J_\mathbf{J}v^K_\mathbf{1}) = \frac{P_0}{2}\; K_{IJK} (w^I_\mathbf{I}w^J_\mathbf{J}v^K_\mathbf{1})\,.
 \end{equation}  
Everything here is real, and we note that this is exactly needed to make \eqref{eq:P_ijk_aas} vanish:
\begin{equation}
    P_{ijk}(w^i_{\mathbf{I}_-}w^j_{\mathbf{J}_-}v^k_{\mathbf{1}_+}) = 0\,, \qquad P_{ijk}(w^i_{\mathbf{I}_+}w^j_{\mathbf{J}_-}v^k_{\mathbf{1}_-}) = 0\,.
\end{equation}
As a consequence, the cubic coupling vanishes for \textit{all} extremal arrangements of the type $(10,5,5)$ and $(6,5,11)$!

Finally, we look at some (super-)extremal arrangements involving also complex structure fields with dimensions $2$ and $3$, more precisely the arrangements $(10,3,3)$ and $(6,3,3)$. These involve two massless axions, and these are flat directions of the potential. As a consequence, the derivative coupling contributes to the effective cubic coupling~\eqref{eq:definition_cprime}, and the mass combination does not vanish. So we must compute quantities of the following type:
\begin{equation}
    d_{ijk} \;y^{i}_\mathbf{I_+}y^{j}_\mathbf{J_-}y^{k}_\mathbf{K_-}. 
\end{equation}
Using eq.~\eqref{eq:fromRealtoComplexContraction} again to transform this expression in terms of the complex fields, we obtain formally eight different terms, but because the field space metric is hermitian, only four remain:
\begin{equation}
    d_{ijk}\; y^{i}_\mathbf{I_+}y^{j}_\mathbf{J_-}y^{k}_\mathbf{K_-} = K_{I \bar J K } y^I_\mathbf{I} \bar y^{\bar J}_\mathbf{J} y^K_\mathbf{K}+K_{\bar I \bar J K } \bar y^{\bar I}_\mathbf{I} \bar y^{\bar J}_\mathbf{J} y^K_\mathbf{K}+K_{I  J \bar K } y^I_\mathbf{I} y^{J}_\mathbf{J} \bar y^{\bar K}_\mathbf{K}+K_{\bar I  J \bar K } \bar y^{\bar I}_\mathbf{I} y^{J}_\mathbf{J} \bar y^{\bar K}_\mathbf{K}.
\end{equation}
We used that $g_{\bar I J} = K_{\bar I J}$. Due to our real parametrisation, this reduces to 
\begin{equation}
    d_{ijk}\; y^{i}_\mathbf{I_+}y^{j}_\mathbf{J_-}y^{k}_\mathbf{K_-} = 4 K_{I J K } y^I_\mathbf{I} y^{J}_\mathbf{J} y^K_\mathbf{K}. 
\end{equation}
The same result holds for derivative couplings involving only saxions.
For the arrangement $(10,3,3)$, we must compute $ K_{IJK}  \left(u_\mathbf{I}^I u_\mathbf{J}^J v_\mathbf{1}^K\right)$. By using that only the K\"ahler potential of the complex structure sector is relevant (because $u_\mathbf{I}$ has only components in this sector) an by using the no-scale identities~\eqref{eq:no_scale_relations}, we find that 
\begin{equation}\label{eq:coupling_1033}
    K_{IJK}  \left(u_\mathbf{I}^I u_\mathbf{J}^J v_\mathbf{1}^K\right) =\frac{1}{2}(u_\mathbf{I}\cdot u_\mathbf{J})\,,
\end{equation}
For arrangements of the type $(6,3,3)$, where the saxion is given by $w_\mathbf{1+}$ instead of $v_\mathbf{1+}$, we find analogously that 
\begin{equation}\label{eq:coupling_633}
    K_{IJK}  \left(u_\mathbf{I}^I u_\mathbf{J}^J w_\mathbf{1}^K\right) = -6(u_\mathbf{I}\cdot u_\mathbf{J})\,.
\end{equation}

It is important that the RHS of eqns.~\eqref{eq:coupling_1033}-\eqref{eq:coupling_633} do not vanish for $\mathbf{I} = \mathbf{J}$, and so there are always non-vanishing (super-)extremal couplings when the eigenvectors $\{u_{\mathbf{I}_\pm}\}$ exist.\footnote{On the other hand, cubic couplings of the type $(10,5,3)$ do vanish.}
Since these eigenvectors only appear when $h^{2,1}\neq 0$, we conclude that the (super-)extremal couplings only vanish in DGKT scenarios when the Calabi-Yau three-fold is rigid, i.e.~when $h^{2,1}=0$. Notice that, as a consequence of $\mathcal{N}=1$ SUSY \cite{Revello:2026opc} (see also \cite{Rong:2018okz}), the cubic couplings for the arrangements  $(10,2,2)$ are proportional to those of the arrangement $(10,3,3)$, and hence they also do not vanish. They involve both non-vanishing potential and derivative couplings, making their computation more involved. Similarly, the cubic couplings for $(6,2,2)$ and $(5,3,2)$ are proportional to those of $(6,3,3)$ and are again non-zero.

\section{Discussion}\label{sec:discussion}
In this paper, we have shown that the spectrum of $\mathcal{N}=1$ supergravity theories in 3d and 4d can easily be determined by computing the eigenvalues of the Hessian of the real superpotential $P$, as they are the conformal dimensions $\Delta$ or $d-\Delta$ ($d=2,3$ for 3d, 4d respectively). In theories with complex scalar fields, this also reveals that the  conformal dimensions between axionic and saxionic partners must differ by $(d-2)$. In spirit, this is much simpler than diagonalising the Hessian of the potential, as this might involve quadratic expressions in the fluxes.\footnote{Although in most scenarios, the flux parameter dependence drops out.}
We emphasise once more that these observations apply also for other theories, in higher dimensions, with more supersymmetry or no supersymmetry, as long as the scalar potential can be written in terms of a function $P$ as in eq.~\eqref{eq:V_ito_P}. In the case of no supersymmetry, this would be referred to as fake supersymmetry.

Then, for supersymmetric setups, we showed that the cubic coupling for an arbitrary arrangement of scalars is entirely dictated by the third derivative of the real superpotential $P$ when there are no derivative couplings for that arrangement, and that the same holds in the presence of derivative couplings for extremal arrangements due to a non-trivial cancellation. 

We then applied these techniques to the supersymmetric DGKT scenario on an arbitrary Calabi-Yau three-fold. We recovered the spectrum as derived in ref.~\cite{Apers:2022tfm}, albeit in a much simpler fashion and with the actual eigenvectors, making extensive use of the no-scale relations of the K\"ahler potential of Calabi-Yau compactifications.
We want to emphasise that integer spectra can only arise when the characteristic polynomial of the eigenvalue problem is a Diophantine polynomial (up to an overall factor). Here, this is a consequence of those no-scale relations, and the simple form of the superpotential: the matrix to diagonalise, i.e.~$\mathcal{P}$  in \eqref{eq:Hessian_P_DGKT}, takes a simple form involving only $K_I$ and $K_{IJ}$ with integer coefficients.
We expect the 3d solutions of refs.~\cite{Arboleya:2024vnp,VanHemelryck:2025qok,Farakos:2025bwf} which were recently discussed in \cite{Revello:2026eqp}, to have a similar structure due to no-scale relations, and leave this for the future.

Then, we computed the extremal cubic couplings coming from the extremal arrangements $(10,5,5)$ and $(6,5,11)$ in the K\"ahler and dilaton sector. We did so by evaluating the third field derivative of the real superpotential, and we found them to always vanish. Together with the results on the spectrum, we interpret this as a clear indication towards the existence of non-trivial structure that becomes much more transparent once DGKT is viewed through the lens of holography. It would be interesting to understand if this can be tied to the existence of additional symmetries, perhaps along the lines of \cite{Bonifacio:2018zex,Apers:2022vfp}. At the same time, we found that some of the super-extremal cubic couplings involving two complex structure moduli and one K\"ahler modulus do not vanish, more precisely for the arrangements $(10,2,2)$ and $(6,2,2)$, with identical scalars of dimension 2. The scalars with dimension 10 and 6 are the universal ones, i.e.~those corresponding to the eigenvectors $v_\mathbf{1}$ and $w_\mathbf{1}$ that are always present and whose form is model-independent. The latter means that, the only way for all (super-)extremal cubic couplings to vanish, is that the Calabi-Yau three-fold has no complex structure moduli and hence is rigid, with $h^{2,1}=0$. 
These general Calabi-Yau results are naturally consistent with the results of refs.~\cite{Bobev:2025yxp,Revello:2026eqp, Revello:2026opc} for DGKT on toroidal orbifolds.
Further, we remark that our analysis was performed for a specific choice of superpotential in which the axions are stabilised at vanishing vev. As shown in ref.~\cite{DeWolfe:2005uu}, DGKT vacua with non-vanishing vevs can be obtained by a field reparametrisation, namely an $\mathrm{SL}(2,\mathbb{Z})$ coordinate transformation in moduli space induced by shifts of the $B$-axions, which may be interpreted as a $B$-twist of the Ramond-Ramond fluxes. Similarly, certain double T-dualities can also be understood as $\mathrm{SL}(2,\mathbb{Z})$ coordinate transformations \cite{Carrasco:2023hta}. Since such transformations come with a diagonal Jacobian, they do not affect our analysis and our results are equally valid for these setups.

Our results also apply to some non-supersymmetric DGKT vacua, related by skew-whiffing to the supersymmetric ones \cite{DeWolfe:2005uu}. These are related by sign changes of the flux parameters that leave the kinetic terms and potential invariant (but not the F-term equations), hence the extremal cubic couplings are the same. However, our techniques do not allow us to conclude the same for the other non-supersymmetric DGKT vacua \cite{DeWolfe:2005uu} that lead to a different (but still integer) spectrum \cite{Quirant:2022fpn}. The fact that the (super-)extremal cubic couplings were found to vanish on the $\mathbb{Z}_3\times\mathbb{Z}_3$ orbifold in ref.~\cite{Bobev:2025yxp} looks encouraging for a generalisation, which would be interesting to investigate in the future.

We also remark that for the supersymmetric DGKT solutions, there are $h^{2,1}$ $C_3$-axions that remain exactly massless, leading to the marginal operators. Non-perturbative D2-instantons lift these moduli and are expected to contribute to their mass schematically as $\exp[-S_{\mathrm{ED2}}]$, which would lead to an exponentially suppressed anomalous dimension as a function of the (large) central charge $c$. Indeed, in the standard DGKT large-flux limit ($e_{\alpha} \sim \bar{e} \rightarrow \infty$), the volumes of (non-collapsing) three-cycles scale as $\bar{e}^{3/4}$, while $g_s\sim \bar{e}^{-3/4}$ and $c\sim \bar{e}^{9/2}$, implying $S_{\mathrm{ED2}}\sim c^{1/3}$.  These will give a sub-leading effect with respect to the polynomially suppressed anomalous dimension of the scalar operators of the K\"ahler + universal CS/dilaton sector, and we expect the constraint to hold also in this case.

Finally, as we already noted in ref.~\cite{Revello:2026eqp}, which also studied 3d solutions, some of the cubic couplings of (super-)extremal arrangements seem non-vanishing when O-planes wrap cycles in different homology classes.
This observation fits well with the results here: after an anti-holomorphic orientifold involution that leads to O6-planes, the Calabi-Yau has $h^{2,1}+1$ orientifold even (and odd) three-cycles that the O6-planes could wrap. The fact that the holographic constraint of ref.~\cite{Bobev:2025yxp} imposes $h^{2,1}=0$, therefore implies indeed that the O-planes should only wrap cycles in the same homology class, and suggests that solutions with $h^{2,1}\neq 0$ belong to the swampland. It would be interesting to further investigate whether there is any deeper principle that connects these two observations.
By contrast, there are other scale-separated solutions, with both integer \cite{Arboleya:2026irl} and non-integer \cite{Farakos:2020phe, Farakos:2025bwf, Miao:2025rgf} spectra, that are not subject to this constraint and in which the O-planes wrap cycles in different homology classes. In those cases, there is no obvious obstruction. Nevertheless, in light of the solutions with \mbox{(super-)extremal} spectra, it would be worthwhile to understand whether additional arguments may rule out such configurations. Resolving this question may further illustrate how holography can expose hidden consistency conditions in string compactifications and lead to an explicit construction of brane duals along the lines of refs.~\cite{Apers:2022vfp,Apers:2025pon,Bedroya:2025ltj,Apers:2026lgi,Arboleya:2026irl}.

\section*{Acknowledgements}
We thank Fien Apers, Ivano Basile, Nikolay Bobev, Joe Conlon, Ulf Danielsson, Suvendu Giri, Vincent Menet, Hynek Paul, Alessandro Tomasiello, Thomas Van Riet and Farah Verbeure for enlightening discussions.
FR acknowledges support
from a junior postdoctoral fellowship of the Fonds Wetenschappelijk Onderzoek (FWO),
project number 12A1Q25N. The work of VVH is supported by Kungliga Fysiografiska s\"allskapet
i Lund.

\bibliographystyle{JHEP}
\bibliography{refs.bib}

\end{document}